\documentclass[9pt,journal,twocolumn,a4paper]{IEEEtran}
\usepackage{times}
\usepackage{amsbsy}
\usepackage{latexsym}
\usepackage{amsmath}
\usepackage[ansinew]{inputenc}
\usepackage[dvips ]{graphicx}
\input epsf
\title{\LARGE Blind Adaptive MIMO
Receivers for Space-Time Block-Coded DS-CDMA Systems in Multipath
Channels Using the Constant Modulus Criterion }
\author{ Rodrigo C. de Lamare and Raimundo Sampaio-Neto   \vspace{-1em}\\
\thanks{\small This work was supported in part by the Brazilian Council for Scientific and
Technological Development (CNPq). R. C. de Lamare is with the
Communications Research Group, Department of Electronics,
University of York, York Y010 5DD, United Kingdom and R.
Sampaio-Neto is with CETUC/PUC-RIO, 22453-900, Rio de Janeiro,
Brazil. Phone: +55-21-31141701 Fax: +55-21-22945748. E-mails:
rcdl500@ohm.york.ac.uk, raimundo@cetuc.puc-rio.br} }

\begin{document}
\maketitle

\begin{abstract}
We propose blind adaptive multi-input multi-output (MIMO) linear
receivers for DS-CDMA systems using multiple transmit antennas and
space-time block codes (STBC) in multipath channels. A space-time
code-constrained constant modulus (CCM) design criterion based on
constrained optimization techniques is considered and recursive
least squares (RLS) adaptive algorithms are developed for
estimating the parameters of the linear receivers. A blind
space-time channel estimation method for MIMO DS-CDMA systems with
STBC based on a subspace approach is also proposed along with an
efficient RLS algorithm. Simulations for a downlink scenario
assess the proposed algorithms in several situations against
existing methods. 
\end{abstract}

\section{Introduction}

\PARstart{T}{he} ever-increasing demand for performance and
capacity in wireless networks has motivated the development of
numerous signal processing and communications techniques for
utilizing these resources efficiently. Recent results on
information theory have shown that higher spectral efficiency
\cite{foschini,telatar} and diversity \cite{alamouti,tarokh1} can
be achieved with multiple antennas at both transmitter and
receiver. Space-time coding (STC) techniques can exploit spatial
and temporal transmit diversity \cite{alamouti,tarokh1,tarokh2}.
The problem of receiver design for DS-CDMA systems using multiple
transmit antennas and STBC has been considered in recent works
\cite{huang}-\cite{grossi}, however, there are still some open
problems. One key issue is the amount of training required by MIMO
channels which motivates the use of blind techniques. In addition,
the existing blind MIMO schemes \cite{Li,reynolds,buzzi,yu,grossi}
are susceptible to the problem of mismatch resulting from
imperfect channel knowledge and this calls for a robust approach.
Prior work on blind techniques for MIMO CDMA systems is limited to
subspace \cite{reynolds} and constrained minimum variance (CMV)
approaches \cite{Li,buzzi,yu,grossi}, which are susceptible to the
problem of signature mismatch. The code-constrained constant
modulus (CCM) approach has demonstrated increased robustness and
better performance than CMV techniques
\cite{xu&tsa}-\cite{ccmstbc-its} for single-antenna systems
although it has not been considered for MIMO systems.

The goal of this work is to propose blind adaptive MIMO receivers
for DS-CDMA systems using multiple transmit antennas and STBC
based on the CCM design in multipath channels. In the proposed
scheme, we exploit the unique structure of the spreading codes and
STBC to derive efficient blind receivers based on the CCM design
and develop computationally efficient RLS algorithms. The proposed
design approach for MIMO receivers requires the knowledge of the
space-time channel. In order to blindly estimate the channel, we
present a subspace approach that exploits the STBC structure
present in the received signal and derive an adaptive RLS type
channel estimator. We also establish the necessary and sufficient
conditions for the channel identifiability of the method. The only
requirement for the receivers is the knowledge of the signature
sequences for the desired user.


\section{Space-time DS-CDMA system model}

\begin{figure*}[t]
\begin{center}
\def\epsfsize#1#2{2.0\columnwidth}
\epsfbox{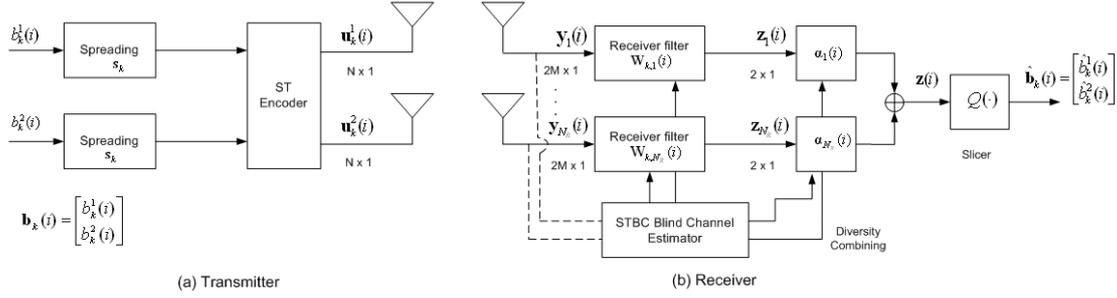} \vspace{-3.5em}\caption{Proposed space-time
system: schematic of the $k$th user (a) Transmitter and (b) Receiver
.}
\end{center}
\end{figure*}
Consider the downlink of a symbol synchronous QPSK DS-CDMA system
shown in Fig. 1 with $K$ users, $N$ chips per symbol, $N_{t}$
antennas at the transmitter, $N_r$ antennas at the receiver and
$L_{p}$ propagation paths. For simplicity, we assume that the
transmitter (Tx) employs only $N_t=2$ antennas and adopts
Alamouti's STBC scheme \cite{alamouti}, although other STBC can
 be used. In this scheme, for $i=1, \ldots, P$ two symbols
$b_{k}(2i-1)$ and $b_{k}(2i)$ are transmitted from Tx1 and Tx2,
respectively, during the $(2i-1)th$ symbol interval and, during
the next symbol interval, $-b_k^*(2i)$ and $b_k^*(2i-1)$ are
transmitted from Tx1 and Tx2, respectively. Each user is assigned
a unique spreading code for each Tx, which may be constructed in
different ways \cite{hochwald,Li}. We assume that the receiver
(Rx) is synchronized with the main path, the delays of the channel
paths are multiples of the chip rate, the channel is constant
during two symbol intervals and the spreading codes are repeated
from symbol to symbol. The received signal at antenna $m$ after
chip-pulse matched filtering and sampling at chip rate over two
consecutive symbols yields the $M$-dimensional received vectors
\begin{equation}
\begin{split}
{\bf r}(2i-1) & = \sum_{k=1}^{K} A_{k}b_{k}(2i-1) {\bf
C}_{k}^1{\bf h}_{m}^1+    A_{k}b_{k}(2i) {\bf C}_{k}^2{\bf
h}_{m}^2 \\ & \quad + \boldsymbol{\eta}_{k,m}(2i-1) + {\bf
n}_m(2i-1)
\\ {\bf r}(2i) & = \sum_{k=1}^{K} A_{k}b_{k}^*(2i-1) {\bf C}_{k}^2{\bf
h}_{m}^2 -     A_{k}b_{k}^*(2i) {\bf C}_{k}^1{\bf
h}_{m}^1 \\ & \quad  + \boldsymbol{\eta}_{k,m}(2i) + {\bf n}_m(2i) \\
&  i=1,~\ldots,P, ~~~~~~~~ m=1,~\ldots, N_r
\end{split}
\end{equation}
where $M=N+L_{p}-1$, ${\bf n}_m(i) = [n_{1}(i)
~\ldots~n_{M}(i)]^{T}$ is the complex Gaussian noise vector with
mean zero and $E[{\bf n}_m(i){\bf n}^{H}_m(i)] = \sigma^{2}{\bf
I}$, where $(.)^{T}$ and $(.)^{H}$ denote transpose and Hermitian
transpose, respectively. The quantity $E[.]$ stands for expected
value and the amplitude of user $k$ is $A_{k}$. The channel vector
for the users' signals transmitted from each transmit antenna
$n_t$ ($n_t=1,2$) and received at the $m$-th receive antenna are
${\bf h}_{m}^{n_t}(i) = [h_{m,0}^{n_t}(i) \ldots
h_{m,L_{p}-1}^{n_t}(i)]^{T}$ and $\boldsymbol{\eta}_{m}(i)$ is the
intersymbol interference at the $m$th receive antenna. The $M
\times L_{p}$ convolution matrix ${\bf C}_{k}^{n_{t}}$ contains
one-chip shifted versions of the signature sequence for user $k$
and each transmit antenna given by ${\bf s}_{k}^{n_{t}} =
[a_{k}^{n_t}(1) \ldots a_{k}^{n_t}(N)]^{T}$ (the reader is
referred to \cite{Li,xu&tsa,delamareccm} for details on the
structure of ${\bf C}_{k}^{n_{t}}$). The received data in (1)
organized into a single $2M \times 1$ vector ${\bf y}_{m}(i) = [
{\bf r}^T(2i-1)~ {\bf r}^T(2i)]^T$ within the $i$th symbol
interval at the $m$th receive antenna is
\begin{equation}
\begin{split}
{\bf y}_{m}(i)  & = \sum_{k=1}^{K} A_{k} b_{k}(2i-1)
\boldsymbol{\mathcal{C}}_k \boldsymbol{g}_m(i) + A_{k}b_{k}(2i)
\bar{\boldsymbol{\mathcal{C}}}_k \boldsymbol{g}^*_m(i) +
\boldsymbol{\eta}_k(i) + {\bf n}(i) \\
& = \sum_{k=1}^{K} {\bf x}_{k}(i) + \bar{\bf x}_{k}(i)  +
\boldsymbol{\eta}_k(i) + {\bf n}(i) \\
\end{split}
\end{equation}
where
\begin{equation}
\boldsymbol{\mathcal{C}}_k = \left[\begin{array}{c c}
{\bf C}_k^1  & {\bf 0}  \\
{\bf 0} & {\bf C}_k^2  \\
 \end{array}\right], \bar{\boldsymbol{\mathcal{C}}}_k = \left[\begin{array}{c c}
{\bf 0} & {\bf C}_k^2  \\
-{\bf C}_k^1 & {\bf 0}   \\
 \end{array}\right], ~{\bf C}_k^{1} ~ {\textrm{ and}} ~ {\bf C}_k^{2} \in
 \boldsymbol{\mathcal{R}}^{M\times L_p} \cite{delamareccm}
\end{equation}
\begin{equation}
\boldsymbol{g}_m(i) = \left[\begin{array}{c} {\bf h}_{k,m}^1 \\
{\bf h}_{k,m}^2\end{array}\right], \boldsymbol{\eta}_k(i) = \left[\begin{array}{c} \boldsymbol{\eta}_{k,1}(2i-1) \\
\boldsymbol{\eta}_{k,2}(2i)\end{array}\right],{\bf n}(i) = \left[\begin{array}{c} {\bf n}_1(2i-1) \\
{\bf n}_2(2i)\end{array}\right]
\end{equation}
The $2M \times 1$ received vectors ${\bf y}_m(i)$ are linearly
combined with the $2M \times 2$ parameter matrix ${\bf
W}_{k,m}(i)$ of user $k$ of the $m$th antenna at the Rx  to
provide the soft estimates
\begin{equation}
{\bf z}_m(i)= {\bf W}^{H}_{k,m}(i){\bf y}_{m}(i) = [ z_{k,m}(i) ~
\bar{z}_{k,m}(i) ]^T
\end{equation}
By collecting the soft estimates ${\bf z}_m(i)$ at the Rx, the
designer can also exploit the spatial diversity at the receiver as
\begin{equation}
{\bf z}(i) = \sum_{m=1}^{N_r} \boldsymbol{\alpha}_{m}(i) {\bf
z}_{m}(i)
\end{equation}
where $\boldsymbol{\alpha}_m(i) = {\rm diag} \big(
\alpha_{m,1}(i),~ \alpha_{m,2}(i) \big)$ are the gains of the
combiner at the receiver, which can be equal leading to Equal Gain
Combining (ECG) or proportional to the channel gains as with
Maximal Ratio Combining (MRC) \cite{rappa}.

\section{Space-time linearly constrained receivers based on the CCM design criterion}

Consider the $2M$-dimensional received vector at the $m$th
receiver ${\bf y}_m(i)$, the $2M\times 2L_{p}$ constraint matrices
$\boldsymbol{\mathcal{C}}_k$ and
$\bar{\boldsymbol{\mathcal{C}}}_k$ that were defined in (3) and
the $2L_{p}\times 1$ space-time channel vector
$\boldsymbol{g}_m(i)$ with the multipath components of the unknown
channels from Tx1 and Tx2 to the $m$th antenna at the receiver.
The space-time linearly constrained receiver design according to
the CCM criterion corresponds to determining an $2M\times 2$ FIR
filter matrix ${\bf W}_{k,m}(i)=\big[ {\bf w}_{k,m}(i), \bar{\bf
w}_{k,m}(i) \big] $ composed of two FIR filters ${\bf w}_{k,m}(i)$
and $\bar{\bf w}_{k,m}(i)$ with dimensions $2M \times 1$. The
filters ${\bf w}_{k,m}(i)$ and $\bar{\bf w}_{k,m}(i)$ provide
estimates of the desired symbols at the $m$th antenna of the
receiver as given by
\begin{equation}
\hat{\bf b}_{k}(i) = {\rm{sgn}} \big( \Re \big[ {\bf
W}_{k,m}^{H}(i){\bf y}_m(i) \big]\big) + j~ {\rm{sgn}}~ \big(\Im
\big[ {\bf W}_{k,m}^{H}(i){\bf y}_m(i) \big]\big)
\end{equation}
where $\rm{sgn}(\cdot)$ is the signum function, $\Re(.)$ selects
the real component, $\Im(.)$ selects the imaginary component and
${\bf W}_{k,m}(i)$ is designed according to the minimization of
the following constant modulus (CM) cost functions
\begin{equation}
\label{costfunction1} J_{CM}({\bf w}_{k,m}(i)) = E\big[(|{\bf
w}_{k,m}^{H}(i){\bf y}_m(i)|^{2}-1)^{2}\big]
\end{equation}
\begin{equation}
\label{costfunction2} J_{CM}(\bar{\bf w}_{k,m}(i)) =
E\big[(|\bar{\bf w}_{k,m}^{H}(i){\bf y}_m(i)|^{2}-1)^{2}\big]
\end{equation}
subject to the set of constraints described by
\begin{equation}
\label{constraints} \boldsymbol{\mathcal{C}}_k^{H}{\bf w}_{k,m}(i)
= \nu ~ \boldsymbol{g}_m(i), ~~~
\bar{\boldsymbol{\mathcal{C}}}_k^{H}\bar{\bf w}_{k,m}(i) = \nu ~
\boldsymbol{g}_m^*(i)
\end{equation}
where $\nu$ is a constant to ensure the convexity of
(\ref{costfunction1}) and (\ref{costfunction2}), which is detailed
in Appendix I along with the convergence properties. The proposed
approach is to consider the design problems in
(\ref{costfunction1}) and (\ref{costfunction2}) via the
optimization of the two filters ${\bf w}_{k,m}(i)$ and $\bar{\bf
w}_{k,m}(i)$ in a simultaneous fashion. The optimization of each
filter aims to suppress the interference and estimate the symbols
transmitted by each transmit antenna. The expressions for the
filters of the space-time CCM linear receiver are derived using
the method of Lagrange multipliers \cite{haykin} and are given by
\begin{equation}
\begin{split}
\label{filter1} {\bf w}_{k,m}(i+1) & = {\bf
R}^{-1}_{k,m}(i)\Big[{\bf d}_{k,m}(i) - \boldsymbol{\mathcal{C}}_k
(\boldsymbol{\mathcal{C}}_k^{H}{\bf
R}^{-1}_{k,m}(i)\boldsymbol{\mathcal{C}}_k)^{-1} \\ & \quad \cdot
\big(\boldsymbol{\mathcal{C}}_k^{H}{\bf R}^{-1}_{k,m}(i) {\bf
d}_{k,m}(i) -\nu~\boldsymbol{g}_m(i)\big)\Big]
\end{split}
\end{equation}
\begin{equation}
\begin{split}
\label{filter2} \bar{\bf w}_{k,m}(i+1) & = \bar{\bf
R}^{-1}_{k,m}(i)\Big[\bar{\bf d}_{k,m}(i) -
\bar{\boldsymbol{\mathcal{C}}_k}
(\bar{\boldsymbol{\mathcal{C}}}_k^{H}\bar{\bf
R}^{-1}_{k,m}(i)\bar{\boldsymbol{\mathcal{C}}}_k)^{-1}  \\ & \quad
\cdot \big( \bar{\boldsymbol{\mathcal{C}}}_k^{H} \bar{\bf
R}^{-1}_{k,m}(i) \bar{\bf d}_{k,m}(i)
-\nu~\boldsymbol{g}_m^*(i)\big)\Big]
\end{split}
\end{equation}
where ${\bf R}_{k,m}(i) = E[|z_{k,m}(i)|^2{\bf y}_m(i){\bf
y}_m^H(i)]$ and $\bar{\bf R}_{k,m}(i) = E[|\bar{z}_{k,m}(i)|^2{\bf
y}_m(i){\bf y}_m^H(i)]$ are correlation matrices, where ${\bf
d}_{k,m}(i)=E[{z}_{k,m}^*(i) {\bf y}_m(i)]$ and  $\bar{\bf
d}_{k,m}(i)=E[\bar{z}_{k,m}^*(i) {\bf y}_m(i)]$ are
cross-correlation vectors, which are originated from the proposed
optimization problem. The expressions (\ref{filter1}) and
(\ref{filter2}) require matrix inversions which lead to a
complexity $O((2M)^3)$. It should also be remarked that
(\ref{filter1}) and (\ref{filter2}) are functions of previous
values of the filter and therefore must be iterated in order to
reach a solution. Since (\ref{filter1}) and (\ref{filter2}) assume
the knowledge of the space-time channel parameters, channel
estimation is required. 

\section{Space-time channel estimation}

In this section, we present a method that exploits the signature
sequences of the desired user and the structure of STBC for blind
channel estimation. Consider the received vector ${\bf y}_m(i)$ at
the $m$th Rx, its associated $2M \times 2M$ covariance matrix
${\bf R}_m=E[{\bf y}_m(i){\bf y}_m^H(i)]$, the space-time $2M
\times 2L_{p}$ constraint matrices $\boldsymbol{\mathcal{C}}_k$
and $\bar{\boldsymbol{\mathcal{ C}}}_k$ given in (6) and the
space-time channel vector $\boldsymbol{g}_m(i)$. From (5) we have
that the $k$th user space-time coded transmitted signals are given
by
\begin{equation}
{\bf x}_k(i)=A_k b_k(2i-1)\boldsymbol{\mathcal{C}}_k
\boldsymbol{g}_m(i), ~~ \bar{\bf x}_k(i)=A_k
b_k(2i)\bar{\boldsymbol{\mathcal{C}}}_k \boldsymbol{g}_m^*(i)
\end{equation}
Let us perform singular value decomposition (SVD) on the
space-time $JM \times JM$ covariance matrix ${\bf R}_m$. 
\begin{equation}
\begin{split}
{\bf R}_m & = \sum_{k=1}^{K}E[{\bf x}_{k}(i){\bf x}_{k}^{H}(i)] +
E[\bar{\bf x}_{k}(i)\bar{\bf x}_{k}^{H}(i)] + E[\boldsymbol{\eta}_k(i)\boldsymbol{\eta}_k^H(i)]+ \sigma^{2}{\bf I} \\
& = [{\bf V}_{s}~ {\bf V}_{n}] \left[\begin{array}{c c}
\boldsymbol{\Lambda}_{s} + \sigma^{2} {\bf I} & {\bf 0} \\ {\bf 0}
& \sigma^{2} {\bf I}
\end{array}\right] [{\bf V}_{s} ~{\bf V}_{n}]^{H}
\end{split}
\end{equation}
where ${\bf V}_{s}$ and ${\bf V}_{n}$ are the signal (that
includes the ISI) and noise subspaces, respectively. Since the
signal and noise subspaces are orthogonal \cite{liu&xu,douko}, we
have the conditions ${\bf V}_{n}^{H} {\bf x}_{k}(i)={\bf
V}_{n}^{H} \boldsymbol{\mathcal{C}}_k \boldsymbol{g}_m(i) = {\bf
0}$ and ${\bf V}_{n}^{H}\bar{\bf x}_{k}(i)={\bf V}_{n}^{H}
\bar{\boldsymbol{\mathcal{C}}}_k \boldsymbol{g}_m^*(i) = {\bf 0}$
and hence we have ${\Omega} =
\boldsymbol{g}_m(i)^{H}\boldsymbol{{\mathcal C}_{k}}^{H}{\bf
V}_{n}{\bf V}_{n}^{H} \boldsymbol{\mathcal{C}}_k
\boldsymbol{g}_m(i)={ 0}$ and $\bar{{\Omega}} =
\boldsymbol{g}_m^{T}(i)\bar{\boldsymbol{\mathcal{C}}}_k^{H}{\bf
V}_{n} {\bf V}_{n}^{H} \bar{\boldsymbol{\mathcal{C}}}_k
\boldsymbol{g}_m^*(i)={ 0}$. From these conditions and taking into
account the conjugate symmetric properties induced by STBC
\cite{Li}, it suffices to consider only ${\Omega}$, which allows
the recovery of $\boldsymbol{g}_m(i)$ as the eigenvector
corresponding to the smallest eigenvalue of the matrix
$\boldsymbol{{\mathcal C}_{k}}^{H}{\bf V}_{n}{\bf V}_{n}^{H}
\boldsymbol{\mathcal{C}}_k$, provided ${\bf V}_{n}$ is known. To
avoid the SVD on ${\bf R}_m$ and overcome the need for determining
the noise subspace rank that is necessary to obtain ${\bf V}_{n}$,
we resort to the following approach.

\textit{Lemma:} Consider the SVD on ${\bf R}_m$ as in (14), then
we have:
\begin{equation}
\lim_{p \rightarrow \infty} ({\bf R}_m/\sigma^{2})^{-p} = {\bf
V}_{n} {\bf V}_{n}^{H}
\end{equation}
\textit{Proof:} Using the decomposition in (14) and since ${\bf I}
+ \boldsymbol{\Lambda}_{s}/\sigma^2$ is a diagonal matrix with
elements strictly greater than unity, by induction we have as
$p\rightarrow \infty$ that $({\bf R}_m/\sigma^{2})^{-p} = {\bf
V}_{n}{\bf V}_{n}^{H}$.

To blindly estimate the space-time channel of user $k$ at the
$m$th antenna of the receiver we propose the optimization:
\begin{equation}
{\hat{\boldsymbol{{g}}}}_{m}(i) = \arg \min_{{{\boldsymbol{{
g}}}}_{m}(i)}~~{{{\boldsymbol{{g}}}_{m}^{H}(i)}
\boldsymbol{{\mathcal C}_{k}}^{H}\hat{\bf
R}^{-p}_m(i)\boldsymbol{{\mathcal
C}_{k}}{{\boldsymbol{{g}}}}_{m}(i)}
\end{equation}
subject to { $||{{\boldsymbol{{g}}}}_{m}(i)||=1$}, where $p$ is an
integer, $\hat{\bf R}_m(i)$ is an estimate of the covariance
matrix ${\bf R}_m(i)$ and whose solution is the eigenvector
corresponding to the minimum eigenvalue of the $JL_{p}\times
JL_{p}$ matrix $\boldsymbol{{\mathcal C}_{k}}^{H}\hat{\bf
R}^{-p}_m(i) \boldsymbol{{\mathcal C}_{k}}$ that can be obtained
using SVD. The performance of the estimator can be improved by
increasing $p$ even though our studies reveal that it suffices to
use powers up to $p=2$ to obtain a good estimate of ${\bf
V}_{n}{\bf V}_{n}^{H}$. For the space-time block coded CCM
receiver design, we employ the matrix ${\bf R}_{k,m}(i)$ instead
of ${\bf R}_m$ to avoid the estimation of both ${\bf R}_m$ and
${\bf R}_{k,m}(i)$, and an equivalence of these matrices is
established in Appendix II. An analysis of the capacity of the
system and the necessary and sufficient conditions for the method
to work is included in Appendix III.

\section{Blind Adaptive RLS Algorithms for Receiver and Channel Parameter Estimation}

In this section we present RLS algorithms for estimating the
parameters of the space-time receiver and channel as described in
Sections III and IV, respectively.

\subsection{RLS Algorithm for CCM Receiver Parameter Estimation}

Considering the expressions obtained for ${\bf w}_{k,m}(i)$ and
$\bar{\bf w}_{k,m}(i)$ in (\ref{filter1}) and (\ref{filter2}),
replacing $E[\cdot]$ with time averages, we can develop an RLS
algorithm through the recursive estimation of the matrices ${\bf
R}^{-1}_{k,m}$, ${\bf R}^{-1}_{k,m}$,
$\boldsymbol{\Gamma}^{-1}_{k,m}(i) =(\boldsymbol{{\mathcal
C}}_{k}^{H}{\bf R}^{-1}_{k,m}(i) \boldsymbol{{\mathcal
C}}_{k})^{-1}$ and $\bar{\boldsymbol{\Gamma}}^{-1}_{k,m}(i)
=(\bar{\boldsymbol{{\mathcal C}}}_{k}^{H} {{\bar{\bf
R}}}^{-1}_{k,m}(i) {\bar{\boldsymbol{\mathcal C}}}_{k})^{-1}$
using the matrix inversion lemma (MIL) and Kalman RLS recursions
\cite{haykin}. The space-time CCM linear receiver estimates are
obtained with
\begin{equation}
\begin{split}
\hat{\bf w}_{k,m}(i+1) & = \hat{\bf R}^{-1}_{k,m}(i)\Big[\hat{\bf
d}_{k,m}(i) - \boldsymbol{\mathcal{C}}_k
\boldsymbol{\Gamma}^{-1}_{k,m}(i)  \\ & \quad \cdot
\Big(\boldsymbol{\mathcal{C}}_k^{H}\hat{\bf R}^{-1}_{k,m}(i)
\hat{\bf d}_{k,m}(i) -\nu~ {\hat{\boldsymbol{g}}}_m(i)\Big)\Big]
\end{split}
\end{equation}
\begin{equation}
\begin{split}
{\hat{\bar{\bf w}}}_{k,m}(i+1) & = {\hat{\bar{\bf
R}}}^{-1}_{k,m}(i)\Big[{\hat{\bar{\bf d}}}_{k,m}(i) -
\bar{\boldsymbol{\mathcal{C}}_k}
{\bar{\boldsymbol{\Gamma}}}^{-1}_{k,m}(i)  \\ & \quad \cdot
 \Big( \bar{\boldsymbol{\mathcal{C}}}_k^{H}
{\hat{\bar{\bf R}}}^{-1}_{k,m}(i) {\hat{\bar{\bf d}}}_{k,m}(i)
-\nu~ {\hat{\boldsymbol{g}}}_m^*(i)\Big)\Big]
\end{split}
\end{equation}
where
\begin{equation}
\hat{\bf d}_{k,m}(i)=\alpha \hat{\bf
d}_{k,m}(i-1)+(1-\alpha)z_{k,m}^{*}(i){\bf y}_m(i)
\end{equation}
\begin{equation}
{\hat{\bar{\bf d}}}_{k,m}(i)=\alpha {\hat{\bar{\bf
d}}}_{k,m}(i-1)+(1-\alpha)\bar{z}_{k,m}^{*}(i){\bf y}_m(i)
\end{equation}
correspond to estimates of ${\bf d}_{k,m}(i)$ and ${{\bar{\bf
d}}}_{k,m}(i)$,respectively. In terms of computational complexity,
the space-time CCM-RLS algorithm requires $O((2M)^{2})$ to
suppress MAI and ISI against $O((2M)^{3})$ required by
(\ref{filter1}) and (\ref{filter2}).

\subsection{RLS Algorithm for Space-Time Channel Estimation}

We develop an RLS algorithm for the estimation of the space-time
channel $\boldsymbol{{g}}_{m}(i)$ at the $m$th receive antenna.
The proposed RLS algorithm avoids the SVD and the matrix inversion
required in (19) via the MIL and a variation of the power method
used in numerical analysis \cite{golub}. Following this approach,
we first compute the inverse of the matrices ${\bf R}^{-1}_{k,m}$
and $\bar{\bf R}^{-1}_{k,m}$ with the MIL, as part of the
space-time receiver design. Then, we construct the matrices
$\boldsymbol{\Gamma}_{k,m}(i)=\boldsymbol{{\mathcal
C}}_{k}^{H}\hat{\bf R}^{-1}_{k,m}(i)\boldsymbol{{\mathcal C}}_{k}$
and ${\bar{\boldsymbol{\Gamma}}}_{k,m}(i)=
\bar{\boldsymbol{{\mathcal C}}}_{k}^{H} {\hat{\bar{\bf
R}}}^{-1}_{k,m}(i) {\bar{\boldsymbol{\mathcal C}}}_{k}$. At this
point, the SVD on the $2L_{p}\times 2L_{p}$ matrices
${{\boldsymbol{\Gamma}}}_{k,m}(i)$ and
${\bar{\boldsymbol{\Gamma}}}_{k,m}(i)$ that requires $O(L_{p}^3)$
is avoided and replaced by a single matrix-vector multiplication,
resulting in the reduction of the corresponding computational
complexity on one order of magnitude and no performance loss. To
estimate the channel and avoid the SVD on
${{\boldsymbol{\Gamma}}}_{k,m}(i)$ and
${\bar{\boldsymbol{\Gamma}}}_{k,m}(i)$, we employ the variant of
the power method introduced in \cite{douko}
\begin{equation}
{\hat{\boldsymbol{g}}}_m (i)= ({\bf I} - \gamma(i)
{\boldsymbol{\Gamma}}_{k,m}(i)) {\hat{\boldsymbol{g}}}_m(i-1)
\end{equation}
where $\gamma(i)=1/tr[{\boldsymbol{\Gamma}}_{k,m}(i)]$ and we make
$\hat{\boldsymbol{g}}_{m}(i)\leftarrow
\hat{\boldsymbol{g}}_{m}(i)/||\hat{\boldsymbol{g}}_{m}(i)||$ to
normalize the channel. It is worth pointing out that due to
certain conjugate symmetric properties induced by STBC, it is
possible to exploit the data record size for estimation purposes
by using both ${\boldsymbol{\Gamma}}_{k,m}(i)$ and
${\bar{\boldsymbol{\Gamma}}}_{k,m}(i)$ and thus the proposed RLS
algorithm computes
\begin{equation}
{\hat{\boldsymbol{g}}}_m (i)= \Big({\bf I} - \vartheta(i)
({\boldsymbol{\Gamma}}_{k,m}(i)+
{\bar{\boldsymbol{\Gamma}}}_{k,m}(i)) \Big)
{\hat{\boldsymbol{g}}}_m(i-1)
\end{equation}
where $\vartheta(i)=1/tr[{\boldsymbol{\Gamma}}_{k,m}(i)+
{\bar{\boldsymbol{\Gamma}}}_{k,m}(i)]$ and the normalization
procedure remains the same. The algorithm in (22) is adopted since
it has a faster convergence than (21) due to the use of more data
samples and spreading codes.

\subsection{Computational Complexity}

We illustrate the computational complexity of the proposed
algorithms and compare them with existing RLS algorithms and
subspace techniques, as shown in Table I. The proposed space-time
CCM-RLS algorithms have a complexity which is quadratic with $N_t
M$, i.e. the number of transmit antennas $N_t$ and proportional to
the processing gain plus the channel order ($M=N+L_p -1$). The
complexities of the trained RLS algorithm and the blind CMV-RLS
\cite{Li} are also quadratic with $N_t M$, whereas the complexity
of the Subspace algorithm of Reynolds \textit{et al.}
\cite{reynolds} is higher due to the subspace computation. The
complexity of the RLS channel estimation algorithm in (21) is
$2(N_t L_p)^2 + N_t L_p$ as it requires the modified power method
on the sum of the $N_t L_p \times N_t L_p$ matrices
${\boldsymbol{\Gamma}}_{k,m}(i)$ and
${\bar{\boldsymbol{\Gamma}}}_{k,m}(i)$. In contrast, the subspace
algorithm of \cite{reynolds} requires an SVD or the use of the
power method on the $N_t M \times N_t M$ matrix ${\bf R}_m$. It
turns out that since $M >> L_p$ in practice, the proposed
algorithm is substantially simpler than the one in
\cite{reynolds}.

\begin{table}[h]
\centering%
\caption{\small Computational complexity of RLS estimation
algorithms per symbol.} {
\begin{tabular}{ll}
\hline
{\small Algorithm} & { \small Multiplications} \\
\hline
\emph{\small \bf STBC-Trained} &  {\small $6(N_t M)^{2}+2N_t M + 2$}   \\
\emph{\bf } &  {\small $$}   \\
\hline
\emph{\small \bf STBC-CCM} &  {\small $5(N_t M)^{2} +3(N_t L_p)^2 + 3N_t^2 L_p M $}   \\
\emph{\bf } &  {\small $ + 4 N_t M + 4 N_t L_p + 2$}   \\
\hline
\emph{\small \bf STBC-CMV} &  {\small $4(N_t M)^{2} + 2N_t^2 L_p M +2(N_t L_p)^2$}   \\
\emph{\bf } &  {\small $ + 3 N_tM + 4N_t L_p + 2$}   \\
\hline
\emph{\small \bf STBC-Subspace} &  {\small $(N_t M)^{2}+N_t^2 L_p M + 3N_t M $}   \\
\emph{\bf } &  {\small $160 N_t MK + 12 N_tM + 30 K^2 +14 K $}   \\
\hline
\end{tabular}
}
\end{table}



\section{Simulations}

In this section we evaluate the bit error rate (BER) performance
of the proposed blind space-time linear receivers based on the CCM
design (STBC-CCM). We also assess the proposed space-time channel
estimation method (STBC-CCM-CE) in terms of mean squared error
(MSE) performance and their corresponding RLS-type adaptive
algorithms. We compare the proposed algorithms with some
previously reported techniques, namely, the constrained minimum
variance (CMV) with a single antenna \cite{xu&tsa} and with STBC
\cite{Li} and the subspace receiver of Wang and Poor without
\cite{wang&poor} and with STBC \cite{reynolds}. The DS-CDMA system
employs randomly generated spreading sequences of length $N=32$,
one or two transmit antennas with the Alamouti STBC
\cite{alamouti} and one or two receive antennas with MRC. The
downlink channels assume that $L_{p}=6$ (upper bound). We use
three-path channels with powers $p_{l,m}^{1,2}$ given by $0$, $-3$
and $-6$ dB, where in each run and for each transmit antenna and
each receive antenna, the second path delay ($\tau_{2}$) is given
by a discrete uniform random variable (d. u. r. v.) between $1$
and $4$ chips and the third path delay is taken from a d. u. r. v.
between $1$ and $5-\tau_{2}$ chips. The sequence of channel
coefficients for each transmit antenna $n_t=1,2$ and each receive
antenna $m=1,2$ is $h_{l,m}^{n_t}(i)=p_{l,m}^{n_t}
\alpha_{l,m}^{n_t}(i)$ ($l=0,1,2,~\ldots$), where
$\alpha_{l,m}^{n_t}(i)$, is obtained with Clarke's model
\cite{rappa}. The phase ambiguity of the blind space-time channel
estimation method in \cite{douko} is eliminated in our simulations
using the phase of { ${\hat{\boldsymbol{g}}}_m (0)$} as a
reference to remove the ambiguity and for fading channels we
assume ideal phase tracking and express the results in terms of
the normalized Doppler frequency $f_{d}T$ (cycles/symbol).
Alternatively, differential modulation can be used to account for
the phase rotations as in \cite{reynolds} or the semi-blind
approach of \cite{Li} adopted.

We evaluate the BER convergence performance of the proposed RLS
algorithms for both receiver and channel parameter estimation in a
scenario where the system has initially $10$ users, the power
distribution among the interferers follows a log-normal
distribution with associated standard deviation of $3$ dB. After
$1500$ symbols, $6$ additional users enter the system and the
power distribution among interferers is loosen to $6$ dB. The
results shown in Fig. 2 indicate that the proposed STBC-CCM
receiver design achieves the best performance among the analyzed
techniques.

\begin{figure}[!htb]
\begin{center}
\def\epsfsize#1#2{1\columnwidth}
\epsfbox{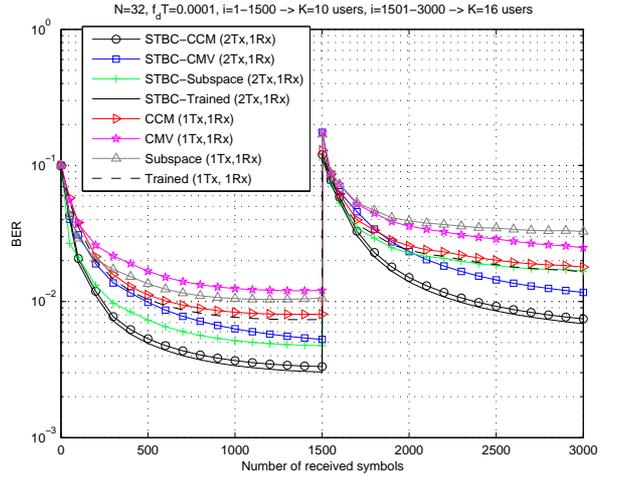} \caption{BER performance versus number of
received symbols in a scenario where users enter the system and
receivers operate at $SNR=E_{b}/N_{0}=15$ dB for the desired user.}
\end{center}
\end{figure}

We assess the channel estimation (CE) RLS algorithms with single
transmit antennas of \cite{wang&poor,douko}, with STBC of Reynolds
{\it et al} \cite{reynolds} and the proposed space-time RLS
channel estimator with STBC given in (21) in terms of MSE between
the actual and the estimated channels using the same dynamic
scenario of the first experiment. The results, shown in Fig 3,
reveal that the proposed space-time channel estimator outperforms
the single-antenna channel estimator because it exploits the
information transmitted by $2$ antennas.

\begin{figure}[!htb]
\begin{center}
\def\epsfsize#1#2{1\columnwidth}
\epsfbox{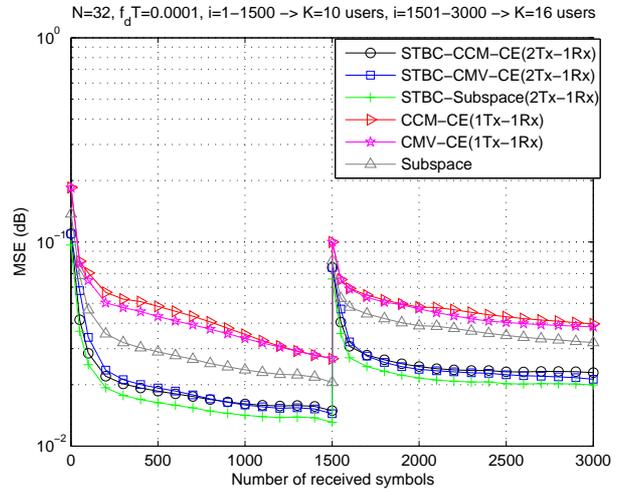} \caption{MSE performance of channel estimation
versus number of received symbols in a dynamic scenario where
receivers operate at $SNR=E_{b}/N_{0}=15$ dB for the desired user.}
\end{center}
\end{figure}

The BER performance versus SNR and number of users is shown in
Fig. 4. We consider data packets of $P=1500$ symbols, $2$ transmit
antennas, $1$ and $2$ receive antennas. We measured the BER after
$200$ independent transmissions. A comparison with previously
reported blind techniques with configurations of $2$ transmit
antennas and $1$ and $2$ receive antennas ((2Tx,1Rx) and
(2Tx,2Rx)) is shown in Fig. 4. The curves illustrate that the
schemes with multiple antennas at the transmitter outperform those
with single-antennas and the capacity of the system is also
increased. With a (2Tx,2Rx) configuration, the diversity is
further exploited and the proposed STBC-CCM achieves a performance
close to the MMSE (also with (2Tx,2Rx)) which assumes the
knowledge of the channel and the noise variance.

\begin{figure}[!htb]
\begin{center}
\def\epsfsize#1#2{1\columnwidth}
\epsfbox{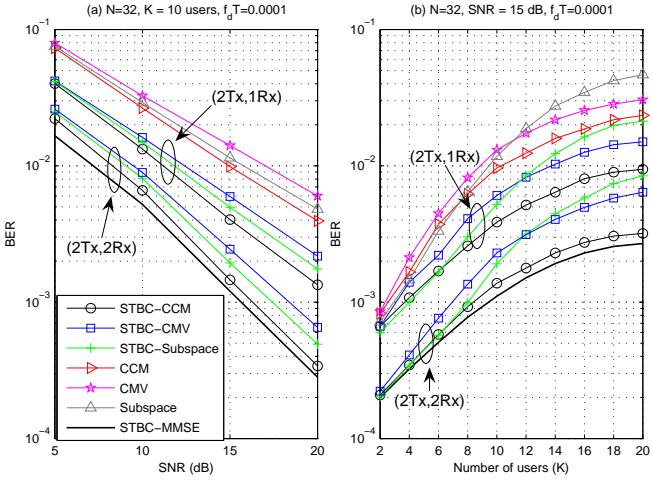} \caption{BER performance versus (a) $SNR$ with
$K=10$ users and (b) number of users (K) at $SNR=E_{b}/N_{0}=15$ dB
.}
\end{center}
\end{figure}

\section{Conclusions}

We presented blind adaptive space-time block-coded linear
receivers for DS-CDMA systems in multipath channels. A CCM design
criterion based on constrained optimization was considered and RLS
algorithms for parameter estimation were developed. We also
derived a blind space-time channel estimation scheme along with an
efficient RLS algorithm. The necessary and sufficient conditions
for the channel identifiability of the proposed method were
established. Simulations for a downlink scenario have shown the
proposed techniques outperform previously reported schemes.

\begin{appendix}

\section{Convergence Conditions}

We present an analysis of the proposed space-time CCM algorithms
and examine their convergence properties and conditions. Let us
consider the cost function expressed as (we will drop the time
index $(i)$ for simplicity)
\begin{equation}
{\mathcal{J}}_{CM} = E \big[(|{\bf w}_{k,m}^H {\bf y}_m |^2 -1
\big] = E \big[|z_{k,m}|^4 - 2 E[|z_{k,m}|^2] + 1
\end{equation}
Let us recall (2) and define ${\bf x} = \sum_{k=1}^{K} {\bf x}_k =
\sum_{k=1}^K A_k b_k(2i-1) {\bf p}_{k,m} $. The received data can
be expressed by ${\bf y}_m = {\bf x} + \bar{\bf x} +
\boldsymbol{\eta} + {\bf n}$. Since the symbols $b_k$ are
independent and identically distributed (i.i.d.) random variables
with zero mean and unit variance, and $b_k$ and ${\bf n}$ are
statistically independent, we have ${\bf R} = {\bf Q} + {\bf T} +
\sigma^2 {\bf I}$, where ${\bf Q} = E[{\bf x} {\bf x}^H], ~~
\textrm{and} ~~ {\bf T} = E[\boldsymbol{\eta} \boldsymbol{\eta}^H
]$. Let us consider user $1$ and antenna $1$ as the desired ones,
let ${\bf w}_1 = {\bf w}$ and define $u_k = A_k {\bf p}_k^H {\bf
w}$ and ${\bf u} = {\bf A}^T {\bf P}^H {\bf w} = [u_1 \ldots
u_K]^T$, where ${\bf A} = \textrm{diag} \big( A_1 \ldots A_K)$,
${\bf P} = [{\bf p}_1, \ldots, {\bf p}_K]$, and ${\bf b} = \big[
{\bf b}_1(2i-1) \ldots {\bf b}_K(2i-1)\big]$. Using the
constraints $\boldsymbol{\mathcal{C}}_1 {\bf w} = \nu
\boldsymbol{g}$ and $\boldsymbol{\mathcal{C}}_1 \bar{\bf w} = \nu
{\boldsymbol{g}}^*$, we have for the desired user the conditions
$u_1 = A_1 {\bf p}_1^H {\bf w} = A_1 \boldsymbol{g}^H$ and $
\boldsymbol{\mathcal{C}}_1^H {\bf w}  = \nu A_1 \boldsymbol{g}^H
{\hat {\boldsymbol{g}}}$. Note that the conditions resemble one
another and only differ by a conjugate term, i.e. $\bar{u}_1 =
u_1^*$. In the absence of noise and neglecting ISI, the (user $1$)
cost function is
\begin{equation}
\begin{split}
{\mathcal{J}}_{CM} ({\bf w}) & = E \big[ ({\bf u}^H {\bf b} {\bf
b}^H {\bf u})^2 \big] -2 E \big[ ({\bf u}{\bf b} {\bf b}^H {\bf u}
) \big] + 1 \\ & = 8 \big( F + \sum_{k=2}^K |u_k|^2 \big)^2 - 4
F^2 \sum_{k=2}^K |u_k|^4 - 4F - 4 \sum_{k=2}^K |u_k|^2 + 1
\end{split}
\end{equation}
where $F = u_1 u_1^* = |u_1|^2 = \nu^2 A_1^2 |\boldsymbol{g}^H
{\hat {\boldsymbol{g}}}|$. Since $\bar{u}_1 = u_1^*$ and $|u_1|^2
= |\bar{u}_1|^2$. The cost functions expressed as above are
equivalent. Thus, it suffices to examine the properties of only
one of them. Let us consider the constraint
$\boldsymbol{\mathcal{C}}_1 {\bf w} = \nu \hat{\boldsymbol{g}}$
and rewrite the cost function ${\mathcal{J}}_{CM} ({\bf w})$ as
\begin{equation}
\begin{split}
{\mathcal{{\mathcal{J}}}}_{CM} ({\bf w})   =
\tilde{{\mathcal{J}}}_{CM} ( {\bf u}) & = 8 (F + {\bf u'}^H {\bf
u}' )^2  - 4 F^2 - 4 \sum_{k=2}^K |u_k|^2 \\ & \quad - 4 F - 4
({\bf u}^{H~'} {\bf u}' )^2 +1
\end{split}
\end{equation}
where ${\bf u}' = \big[ u_2 \ldots u_K]^T =
\boldsymbol{\mathcal{G}} {\bf w}$, $\boldsymbol{\mathcal{G}} =
{\bf A'}^H {\bf P'}^H$, ${\bf P}' = \big[{\bf p}_2 \ldots {\bf
p}_K \big]$ and ${\bf A} = \textrm{diag} \big( A_2 \ldots A_K
\big)$. To evaluate the convexity of $\tilde{{\mathcal{J}}}_{CM}
({\bf u}')$, we compute its Hessian (${\bf H}$) using the
differentiation rule ${\bf H} = \frac{\partial
\tilde{{\mathcal{J}}}_{CM} ({\bf u}') }{\partial {\bf u'}^H}
\frac{ \partial ( \tilde{{\mathcal{J}}}_{CM} ({\bf u}'))}
{\partial {\bf u}'}$ which yields:
\begin{equation}
{\bf H} = \big[ 16(F-\frac{1}{4}) {\bf I} + 16 {\bf u'}^H{\bf u}'
{\bf I} + 16 {\bf u}{\bf u'}^H - 16 \textrm{diag} \big(|u_2|^2
\ldots |u_k|^2\big)
\end{equation}
Specifically, ${\bf H}$ is positive definite if ${\bf a}^H{\bf H}
{\bf a} > 0$ for all nonzero ${\bf a} \in
\boldsymbol{\mathbf{C}}^{K-1 \times K-1}$. The second, third and
fourth terms of ${\bf H}$ are positive definite matrices, whereas
the first term provides the following condition for convexity
\begin{equation}
\nu A_1^2 | \boldsymbol{g}^H \hat{\boldsymbol{g}}| \geq
\frac{1}{4}
\end{equation}
Since ${\bf u}' = \boldsymbol{\mathcal{G}} {\bf w}$ is a linear
function of ${\bf w}$ then $\tilde{{\mathcal{J}}}_{CM} ({\bf u}')$
being a convex function of ${\bf u}'$ implies that $
{\mathcal{J}}_{CM} ({\bf w}) = \tilde{{\mathcal{J}}}_{CM}
(\boldsymbol{\mathcal{G}} {\bf w}) ~\textrm{and}~
{\bar{\mathcal{J}}}_{CM} (\bar{\bf w}) =
\tilde{{\mathcal{J}}}_{CM} (\bar{\boldsymbol{\mathcal{G}}}
\bar{\bf w})$ are convex function of ${\bf w}$ and $\bar{\bf w}$,
respectively, where $\bar{\boldsymbol{\mathcal{G}}} = {\bf A'}^H
\bar{\bf P'}^H$ and $\bar{\bf P}' = \big[\bar{\bf p}_2 \ldots
\bar{\bf p}_K \big]$. As the extrema of the cost functions can be
considered for small noise variance $\sigma^2$ a slight
perturbation of the noise free case, the cost functions will also
be convex for small $\sigma^2$ when $\nu A_1^2 | \boldsymbol{g}^H
\hat{\boldsymbol{g}}| \geq \frac{1}{4}$. If we assume ideal
channel estimation, i.e. $| \boldsymbol{g}^H
\hat{\boldsymbol{g}}|= 1$, and $\nu=1$, the condition will
collapse to $|A_1|^2 \geq \frac{1}{4}$, which corroborates
previous results with the constant modulus algorithms
\cite{xu&liu}. In the case of larger values of $\sigma^2$, the
designer should adjust $\nu$ in order to enforce the convexity of
the cost functions in (8) and (9).

\section{Space-Time Channel and Parameter Estimation with ${\bf
R}_{k,m}$}

We discuss here the the suitability of the matrix ${\bf R}_{k,m}$,
that arises from the space-time CCM design method, for use in the
space-time channel estimator. From the analysis in Appendix I for
the linear receiver, we have for an ideal and asymptotic case that
(we consider receive antenna $1$ and user $1$ for simplicity)
$u_{k} = (A_{1}{\bf p}_{k}^{H}){\bf w}_{1} \approx 0, \textrm{for}
~ \textrm{for}~ k=2,\ldots,K$. Then, we have that ${\bf
w}^{H}_{1}{\bf y} \approx A_{1}b_{1}(2i-1){\bf w}^{H}_{1}{\bf
p}_{1} + {\bf w}^{H}_{1}{\bf n}$, and $|{\bf w}^{H}_{1}{\bf
y}|^{2} \approx A_{1}^{2}|{\bf w}_{1}^{H}{\bf p}_{1}|^{2} +
A_{1}b_{1}(2i-1)({\bf w}_{1}^H{\bf p}_{1}){\bf n}^{H}{\bf w}_{1} +
A_{1}b_{1}^{*}(2i-1)({\bf p}_{1}^{H}{\bf w}_{1}){\bf
w}_{1}^{H}{\bf n} + {\bf w}^{H}_{1}{\bf n}{\bf n}^{H}{\bf w}_{1}$.
Therefore, we have for the desired user :
\begin{equation}
\begin{split}
{\bf R}_{1} & =E[|{\bf w}^{H}_{1}{\bf y}|^{2}{\bf y}{\bf y}^{H}]
 \cong {\beta} {\bf R} + \tilde{\bf N}
\end{split}
\end{equation}
where ${\bf R} = {\bf Q} +\sigma^2{\bf I}$, ${\bf Q} = E[{\bf
x}{\bf x}^{H}] = \sum_{k=1}^{K}|A_{k}|^{2}{\bf p}_{k}{\bf
p}_{k}^{H}$, the scalar factor is ${{\beta}} = A_1^2 \big( |{\bf
w}_{1}^{H}{\bf p}_{1} |^2 + \sigma^2 \big)$ and the noise-like
term is ${\tilde{{\bf N}}}  = A_1^2 \sigma^2 \big[ \big( {\bf
w}_{1}^{H}{\bf p}_{1} ({{\bf w}}_1 {{\bf p}}^H_1) + \sigma ({{\bf
p}}^H_1 {{\bf w}}_1) (  {{\bf p}}_1{{\bf w}}_1^H)  + \big( \big[
\textrm{diag} (|{{\bf w}}_1|^2 \ldots |{{\bf w}}_{2M}|^2 ) + {{\bf
w}} {{\bf w}}^H \big] - {{\bf w}}^H {\bf w} {\bf I} \big) \big]$.

The conditions presented show that ${\bf R}_k$ for a general user
$k$ can be approximated by ${\bf R}$ multiplied by a scalar factor
${\beta}$ plus a noise-like term $\tilde{\bf N}$, that for
sufficient signal-to-noise ratio (SNR) values has an insignificant
contribution. The same analysis applies to $\bar{\bf R} =
E\big[|\bar{\bf w}^H_1 {\bf y}|^2 {\bf y}{\bf y}^H \big]$, which
is given by
\begin{equation}
\bar{\bf R}_1 \cong \bar{\boldsymbol{\beta}} {\bf R} +
\tilde{\bar{\bf N}}
\end{equation}
where $\bar{{\beta}} = A_1^2 \big( |\bar{\bf w}^H_1 \bar{\bf p}_1
|^2 + \sigma^2 \big)$ and ${\tilde{\bar{\bf N}}}  = A_1^2 \sigma^2
\big[ \big( {\bar{\bf w}}_1^H {\bar{\bf p}}_1 ({\bar{\bf w}}_1
{\bar{\bf p}}^H_1) + \sigma ({\bar{\bf p}}^H_1 {\bar{\bf w}}_1) (
{\bar{\bf p}}_1{\bar{\bf w}}_1^H)  + \big( \big[ \textrm{diag}
(|{\bar{\bf w}}_1|^2 \ldots |{\bar{\bf w}}_{2M}|^2 ) + {\bar{\bf
w}} {\bar{\bf w}}^H \big] - {\bar{\bf w}}^H {\bf w} {\bf I} \big)
\big]$. An interesting interpretation of this behavior is the fact
that when the symbol estimates $z_{k}={\bf w}_{k}^{H}{\bf y}$ and
$\bar{z}_{k}=\bar{\bf w}_{k}^{H}{\bf y}$ are reliable and the cost
functions in (8) and (9) are sufficiently small, then
$|z_{k}|^{2}$ and $|\bar{z}_{k}|^{2}$ have small variations around
unity, yielding the approximation $E[|z_{k}|^{2}{\bf y}{\bf
y}^{H}] = E[{\bf y}{\bf y}^{H}] + E[(|z_{k}^{2} - 1){\bf y}{\bf
y}^{H}] \cong E[{\bf y}{\bf y}^{H}]={\bf R}$ and $
E[|\bar{z}_{k}|^{2}{\bf y}{\bf y}^{H}] = E[{\bf y}{\bf y}^{H}] +
E[(|\bar{z}_{k}^{2} - 1){\bf y}{\bf y}^{H}] \cong E[{\bf y}{\bf
y}^{H}]={\bf R}$. Therefore, we can employ ${\bf R}_{k}$ in lieu
of ${\bf R}$, since the properties of ${\bf R}$ studied for the
proposed space-time channel estimation method hold for ${\bf
R}_{k}$.

\section{Identifiability of Space-Time Channels and Consistency of the Estimates}

We develop an expression of the capacity of the space-time system
and discuss necessary and sufficient conditions for the
identifiability of space-time channels and the consistency of the
estimates. Let $q_{s}$ and $q_{n}$ denote the signal and the noise
subspace ranks, respectively. The matrix ${\bf V}_{n}^{H}
\boldsymbol{{\mathcal C}_{k}}$ of dimensions $r_{n} \times JL_{p}$
will be used for our analysis. If the noise subspace ${\bf V}_{n}$
is the exact subspace then, due to ${\bf V}_{n}^{H}
\boldsymbol{{\mathcal C}_{k}}\boldsymbol{{g}_{k}}= {\bf 0}$, we
can verify that the column rank of ${\bf V}_{n}^{H}
\boldsymbol{{\mathcal C}_{k}}$ can at most be $JL_{p}-1$. In order
to have a unique solution (times a phase ambiguity) the column
rank of ${\bf V}_{n}^{H} \boldsymbol{{\mathcal C}_{k}}$ must be
exactly equal to $N_TL_{p}-1$. Since a column rank of a square
matrix is equal to its row rank, a necessary condition for a row
rank equal to $N_TL_{p}-1$ is to have at least $N_TL_{p}-1$ rows,
i.e. $q_{n} \geq N_TL_{p}-1$. Since $q_{s}+q_{n}=N_TM$ this yields
\begin{equation}
q_{s} \leq N_TM - N_TL_{p} +1
\end{equation}

Consider now the signal subspace rank $q_s$ and assume
symbol-by-symbol estimation in a synchronous downlink system. The
number of columns of ${\bf V}_{s}$ (which is an orthonormal basis
for ${\bf x}_{k}$) is $q_{s}$ and is composed by the effective
spatial signatures of all $K$ users transmitted by the $N_T$
antennas, that corresponds to a matrix with size $N_TM \times K$,
and ISI. The ISI corresponds to a matrix with dimensions
$N_T(L_{p}-1) \times K$ since for our transmit diversity
configuration we have $N_T$ independent multipath channels with a
maximum of $L_p$ paths each. Assuming that $K <N_T N$, we have
that if $K + 2 \min \{N_T( L_{p}-1), K \} \leq N_TM - N_TL_{p} +1
= N_T(N-1) + 1$ then the necessary condition on $q_{s}$ is
satisfied and an upper bound for the maximum load of the system
with $N_T$ transmit antennas is
\begin{equation}
K  \leq N_T \Big[N - \frac{(N_T-1)}{N_T} - 2 \min \Big\{
{\Big(\frac{N}{3} - \frac{(N_T-1)}{3 N_T}\Big)}, L_{p}-1 \Big\}
\Big]
\end{equation}
The above result indicates an increase in the system capacity as
compared to the result $K  \leq N - 2 \min \{N/3 , L_{p}-1 \}$ for
a single antenna system reported in \cite{douko}. In order to
increase the capacity, a designer should choose $N_T \geq 2$ .

Now, we are interested in establishing the sufficient conditions
for the identifiability of $\boldsymbol{{g}_{k}}$. We will follow
the approach reported in \cite{liu&xu}. Let us first consider the
effective space-time signatures $\boldsymbol{{\mathcal
C}_{k}}\boldsymbol{{g}_{k}}$ and $\boldsymbol{\bar{\mathcal
C}_{k}}\boldsymbol{{g}_{k}^*}$, and rewrite them as
\begin{equation}
\boldsymbol{{\mathcal C}_{k}} \boldsymbol{{g}_{k}} = {\bf v}_k =
\boldsymbol{{\mathcal X}} {\bf e}_k / A_k, ~~~~
\boldsymbol{\bar{\mathcal C}_{k}} \boldsymbol{{g}_{k}^*} =
\bar{\bf v}_k = \boldsymbol{\bar{\mathcal X}} {\bf e}_k / A_k,
\label{rew}
\end{equation}
where ${\bf e}_k = [\underbrace{0,~ \ldots, ~ 0,}_{k-1} ~ 1, ~
\underbrace{0,~ \ldots, ~ 0}_{2M-k}]^T$. From the above we have
${\rm dim}  \{{\rm range}(\boldsymbol{{\mathcal C}_{k}}) \cap {\rm
range}(\boldsymbol{{\mathcal X}}) \} = 1, ~ {\rm dim}  \{{\rm
range}(\boldsymbol{\bar{\mathcal C}_{k}}) \cap {\rm
range}(\boldsymbol{\bar{\mathcal X}}) \} = 1$ where ${\rm dim}\{
\cdot \}$ stands for the dimension of a subspace and
$\boldsymbol{{\mathcal X}} = \sum_{k=1}^{K}E[{\bf x}_{k}(i){\bf
x}_{k}^{H}(i)]$ and $\boldsymbol{\bar{\mathcal X}} =
\sum_{k=1}^{K}E[\bar{\bf x}_{k}(i)\bar{\bf x}_{k}^{H}(i)]$. The
following theorem establishes the identifiability result for the
proposed method.

\textit{ Theorem}: Let $\boldsymbol{{\mathcal C}_{k}}$,
$\boldsymbol{{\mathcal X}}$, $\boldsymbol{\bar{\mathcal C}_{k}}$
and $\boldsymbol{\bar{\mathcal X}}$ be full-column rank matrices
and $\alpha$ be an arbitrary scalar. Then the equations
\begin{equation}
\begin{split}
{\bf V}_{n}^{H} \boldsymbol{{\mathcal C}_{k}}
\boldsymbol{\tilde{g}}  = {\bf 0}, ~~~~  {\bf V}_{n}^{H}
\boldsymbol{\bar{\mathcal C}_{k}} \boldsymbol{\tilde{g}^*}  = {\bf
0}, \label{rel}
\end{split}
\end{equation}
have a nontrivial solution other than $\alpha
\boldsymbol{{g}_{k}}$ if and only if the following condition
holds:

\textit{ Condition}: There exists the vectors
$\boldsymbol{\tilde{g}} \neq \alpha \boldsymbol{{g}_{k}}$, ${\bf
q}_k$ and $\bar{\bf q}_k$ such that $\boldsymbol{{\mathcal
C}_{k}}\boldsymbol{\tilde{g}}  = \boldsymbol{{\mathcal X}} {\bf
q}_k, ~~~~ \boldsymbol{\bar{\mathcal C}_{k}}\boldsymbol{\tilde{g}}
= \boldsymbol{\bar{\mathcal X}} \bar{\bf q}_k$ which are
equivalent to ${\rm dim} \{{\rm range}(\boldsymbol{{\mathcal
C}_{k}}) \cap {\rm range}(\boldsymbol{{\mathcal X}}) \} = 1,~~~~
{\rm dim} \{{\rm range}(\boldsymbol{\bar{\mathcal C}_{k}})  \cap
{\rm range}(\boldsymbol{\bar{\mathcal X}}) \} = 1$ respectively.

{\textit Proof}: The equations in (\ref{rel}) can be combined
together and written as follows $ {\bf V}_{n}^{H}
\boldsymbol{{\mathcal C}_{k}} \boldsymbol{\tilde{g}} + {\bf
V}_{n}^{H} \boldsymbol{\bar{\mathcal C}_{k}}
\boldsymbol{\tilde{g}^*}= {\bf 0} $. Let the latter condition be
satisfied with some $\boldsymbol{\tilde{g}} \neq \alpha
\boldsymbol{{g}_{k}}$. As the matrix $\big[
 \boldsymbol{{\mathcal X}} ~
\boldsymbol{\bar{\mathcal X}} \big] $ is full-column rank, the
relation $ {\bf V}_{n}^{H} \boldsymbol{{\mathcal C}_{k}}
\boldsymbol{\tilde{g}} + {\bf V}_{n}^{H} \boldsymbol{\bar{\mathcal
C}_{k}} \boldsymbol{\tilde{g}^*}= {\bf 0} $ directly yields $
\big[(\boldsymbol{{\mathcal C}_{k}} \boldsymbol{\tilde{g}})^T ~
(\boldsymbol{\bar{\mathcal C}_{k}} \boldsymbol{\tilde{g}^*})^T
 \big] \in  {\rm range} \big( \boldsymbol{{\mathcal X}} ~
\boldsymbol{\bar{\mathcal X}}\big)$. Hence, from the above we have
$\boldsymbol{{\mathcal C}_{k}}\boldsymbol{\tilde{g}}  =
\boldsymbol{{\mathcal X}} {\bf q}_k, ~~~~
\boldsymbol{\bar{\mathcal C}_{k}}\boldsymbol{\tilde{g}}  =
\boldsymbol{\bar{\mathcal X}} \bar{\bf q}_k$. The relations ${\rm
dim} \{{\rm range}(\boldsymbol{{\mathcal C}_{k}}) \cap {\rm
range}(\boldsymbol{{\mathcal X}}) \} = 1$ and ${\rm dim} \{{\rm
range}(\boldsymbol{\bar{\mathcal C}_{k}})  \cap {\rm
range}(\boldsymbol{\bar{\mathcal X}}) \} = 1$ follows from
$\boldsymbol{{\mathcal C}_{k}}\boldsymbol{\tilde{g}} =
\boldsymbol{{\mathcal X}} {\bf q}_k, ~~~~
\boldsymbol{\bar{\mathcal C}_{k}}\boldsymbol{\tilde{g}} =
\boldsymbol{\bar{\mathcal X}} \bar{\bf q}_k$ along with
(\ref{rew}) and the fact that the vectors $\boldsymbol{{g}_{k}}$
and $\boldsymbol{\tilde{g}}$ are linearly independent. This proves
the condition in $\boldsymbol{{\mathcal
C}_{k}}\boldsymbol{\tilde{g}}  = \boldsymbol{{\mathcal X}} {\bf
q}_k, ~~~~ \boldsymbol{\bar{\mathcal C}_{k}}\boldsymbol{\tilde{g}}
= \boldsymbol{\bar{\mathcal X}} \bar{\bf q}_k$. Now, to prove the
sufficiency part, let us assume that the condition
$\boldsymbol{{\mathcal C}_{k}}\boldsymbol{\tilde{g}}  =
\boldsymbol{{\mathcal X}} {\bf q}_k, ~~~~
\boldsymbol{\bar{\mathcal C}_{k}}\boldsymbol{\tilde{g}} =
\boldsymbol{\bar{\mathcal X}} \bar{\bf q}_k$ holds. From the
latter, we have $\boldsymbol{{\mathcal
C}_{k}}\boldsymbol{\tilde{g}}  = \boldsymbol{{\mathcal X}} {\bf
q}_k$ and $\boldsymbol{\bar{\mathcal C}_{k}}\boldsymbol{\tilde{g}}
= \boldsymbol{\bar{\mathcal X}} \bar{\bf q}_k$ and $
\big[(\boldsymbol{{\mathcal C}_{k}} \boldsymbol{\tilde{g}})^T ~
(\boldsymbol{\bar{\mathcal C}_{k}} \boldsymbol{\tilde{g}^*})^T
 \big] \in  {\rm range} \big( \boldsymbol{{\mathcal X}} ~
\boldsymbol{\bar{\mathcal X}}\big)$. It follows that
$\boldsymbol{\tilde{g}} \neq \alpha \boldsymbol{{g}_{k}}$ is a
solution to $ {\bf V}_{n}^{H} \boldsymbol{{\mathcal C}_{k}}
\boldsymbol{\tilde{g}} + {\bf V}_{n}^{H} \boldsymbol{\bar{\mathcal
C}_{k}} \boldsymbol{\tilde{g}^*}= {\bf 0} $. This completes the
proof.

\end{appendix}

\end{document}